\documentstyle[aps,epsf]{revtex}  

\newcommand{\tpi}{{\tilde{\pi}}}
\newcommand{\be}{\begin{equation}}
\newcommand{\ee}{\end{equation}}
\newcommand{\ber}{\begin{eqnarray}}
\newcommand{\eer}{\end{eqnarray}}
\newcommand{\bers}{\begin{eqnarray*}}
\newcommand{\eers}{\end{eqnarray*}}
\begin{document}        

\baselineskip 14pt
\title{
{\Large \bf{Flavor Changing Decays of $\Upsilon$ and $J/\psi$\\}}
}
\author{Alakabha Datta}
\address{
{\it Department of Physics  \\}
{\it  University of Toronto, \\}
{\it Toronto, Ontario, Canada M5S 1A7\\}
}
%
\maketitle              

\begin{abstract}        
We study flavor changing processes
$\Upsilon \rightarrow B/{\overline B} X_s$ and 
$J/\psi \rightarrow D/{\overline D} X_u$ in the B factories and 
the Tau-Charm factories. In the standard model, these processes are 
predicted to be 
unobservable, so they serve as a probe of the new physics.
We first perform a model independent analysis, then examine the predictions of 
models; such as top color models and MSSM with R-parity
violation ; for the branching ratios of
$\Upsilon \rightarrow B/{\overline B} X_s$ and $J/\psi \rightarrow D/{\overline D} X_u$ 
. We find that these branching ratios 
 could be as large as $10^{-6}$ and $10^{-5}$ in the presence of 
new physics.     
\
\end{abstract}   	

\section{Introduction}               
The possibility of observing large CP violating asymmetries in the
decay of $B$ mesons has motivated the construction of high luminosity $B$
factories at several of the world's high energy physics laboratories. 
These $B$ factories will be producing roughly about $10^8$ Upsilons.
Meanwhile BES has already accumulated $9 \times 10^6$ $J/\psi$
and plans to increase the number to $5\times 10^7$ in the near future.
An interesting question, that we  investigate in this
paper, is whether the large sample of the $\Upsilon$ and the $J/\psi$ can be used to
probe flavor changing processes in the decays of $\Upsilon$ and $J/\psi$.
 In particular we  look at the
flavor changing processes $\Upsilon \to B/{\overline B} X_s$
and $J /\psi \to D/{\overline D} X_u$,   from the
 underlying 
$b \to s$ and $c \to u $ quark transitions. For the quarkonium 
system, these flavor changing processes 
are expected to be much smaller than in the case of decays of
 the $B$ or the $D$
meson
because of the larger decay widths of the bottonium and the charmonium systems
which decay via the strong interactions. Indeed the standard model contributions
to $\Upsilon \to B/{\overline B} X_s$
and $J /\psi \to D/{\overline D} X_u$ are tiny. However, new physics may
enhance the branching ratios for these processes. Whether 
this enhancement maybe sufficient for these
processes to be observable in the next round of experiments is the 
subject of this work.

 Non leptonic decays of
heavy quarkonium systems can be more reliably calculated
 than the non leptonic decays of the heavy mesons. A consistent 
and systematic
formalism to handle heavy quarkonium decays is 
available in NRQCD \cite{lepage} which
is missing for the heavy mesons. As in the meson system 
\cite{datta1}        
it is more fruitful to concentrate 
on quasi-inclusive  processes like
$\Upsilon \to B/{\overline B} X_s$
and $J /\psi \to D/{\overline D} X_u$ because they 
can be calculated with  less theoretical uncertainty and have larger branching ratios
 than the purely
exclusive quarkonium  non leptonic decays. The branching ratios of
exclusive flavor changing non leptonic decays
of $\Upsilon$ and $J/\psi$ in the standard model
 have been calculated and found to be  very small \cite{indianguys}.

We begin with a model independent description of the processes
$\Upsilon \to B/{\overline B} X_s$
and $J /\psi \to D/{\overline D} X_u$. In the standard model
these decays can proceed through tree and penguin processes.
 For new
physics contribution to these processes 
we concentrate on four quark operators
of the type ${\overline s}b{\overline b} b$ and  
${\overline u}c{\overline c} c $. 
We choose the currents in the four quark operators to be scalars and so
these operators may arise through the exchange of a heavy scalar 
for e.g a Higgs or 
a leptoquark in some model of new physics. These four quark operators, 
at the one loop level, generate effective
${\overline s}b \{g,\gamma,Z \}$ and
${\overline u}c \{g,\gamma,Z \}$ vertices
 which 
would effect the flavor changing decays of
the $B$ and the $D$ mesons. The effective vertices for an on shell $g$ and $\gamma$
vanish and so there is no contribution to $b \to s \gamma$ or $c \to u \gamma$. We can 
however put constraints on these operators by considering the processes
$b \to s l^+ l^{-}$ and $c \to u l^+ l^{-}$.  
The constrained operators can
then be used to 
calculate the branching ratios for $\Upsilon \to B/ {\overline B} X_s$
and $J /\psi \to D/ {\overline D} X_u$. 

We then consider some models that may generate the kind of 
four quark operators described above. A few examples of models
where these operators can be generated are
top color models and  MSSM with R parity violation. In some cases 
constraints on
the parameters that appear in the prediction for the branching ratios for
$\Upsilon \to B/{\overline B} X_s$
and $J /\psi \to D/{\overline D} X_u$ are already available. In other cases the parameters 
are constrained, as in our  model independent analysis, from the processes
$ b \to s l^+ l^{-}$ and $c \to u l^+ l^{-}$.

\section{Effective Hamiltonian.}
In the Standard Model (SM) 
the amplitudes for hadronic $\Upsilon$ decays of the type $b\bar{b}\to s
\bar{b} + \bar{s}b$ 
are generated by the an effective 
Hamiltonian \cite{Reina,buras}
To the standard model contribution we add higher dimensional 
four quark operators generated by physics beyond the standard model\cite{DOPZ}.
In this paper, we consider the four quark operators with two scalar
currents.
\ber
L_{new} & = &\frac{R_1}{\Lambda^2}{\overline s}(1-\gamma^5)b{\overline b} 
(1+\gamma^5) b +
\frac{R_2}{\Lambda^2} {\overline s}(1+\gamma^5)b{\overline b}
(1-\gamma^5) b +h.c . \
\eer
\noindent
The four quark operators in $L_{new}$ are the product of two scalar currents.
In Eq. (1)  $\Lambda$ represents the new physics scale and $R_1$ and $R_2$ are two free
parameters which describe the strength of the contribution of the underlying new physics
to the effective operators. In our analysis we will only keep dimension 
six operators suppressed by $1/\Lambda^2$ and neglect all higher 
dimension operators. The details of the matrix elements for the processes
 $\Upsilon \to B/ {\overline B} X_s$
and $J /\psi \to D/ {\overline D} X_u$ can be found in \cite{DOPZ} 
 
\section{Low Energy Constraints and Models}
The lagrangian $L_{new}$ generates, at one loop level, the effective
${\overline s}b\gamma ^*$,
${\overline s}bg^*$,${\overline s}b Z$ vertices, where
 $\gamma^*$ and $g^*$ indicate an off shell photon and a gluon.
Similar vertices involving $c \to u$ transitions 
 are generated in the charmonium sector also. These 
vertices, with a $\gamma$ and $Z$, will contribute to $b \to s l^+l^-$ and
$c \to u l^+l^-$. Note there is no contribution to $b \to s \gamma$. The vertex
$b \to s g^*$ can give rise to the process $b \to s q {\overline q}$ which 
will contribute to non-leptonic $B$ decays. We expect the constraints from
$b \to s l^+l^-$ to be better than from non-leptonic $B$ decays 
because of the theoretical uncertainties in calculating non-leptonic decays
The additional contribution
to the effective Hamiltonian for  $b \to s l^+ l^-$ can be written as
\ber
\delta H_{b \to s l^+ l^-} & = & -\frac{e^2}{16\pi^2}\frac{e_b}{\Lambda^2}
\int^1_0 dx 8x(1-x)\log \left(\frac{\Lambda^2}{B^2}\right)
\left[R_1{\overline s}\gamma^{\mu}b_L{\overline l}\gamma_{\mu}l+   
R_2{\overline s}\gamma^{\mu}b_R{\overline l}\gamma_{\mu}l\right]\
\eer   
which has to be added to the standard model contribution \cite{buras}.
Similar results can also be written for the charm sector.

Now we look at various models that can give rise to 
$L_{new}$ given in Eq. 1.
As a first example we consider a recent version of
 top color models\cite{hill}. In such models the 
top quark participates in a new strong interaction which is broken at 
some high energy 
scale $\Lambda$. The strong interaction, though not confining, leads to the 
formation of a top condensate $<{\overline t}_Lt_R>$ resulting in a large
 dynamical mass for the top quark. The scale $\Lambda$ is chosen to 
be of the order of a TeV to avoid naturalness problem which implies that the 
electroweak symmetry cannot be broken solely by the top condensate.
In the low energy sector of the theory, scalar bound states are formed that 
couple strongly to the $b$ quark \cite{kominis,burdman}
\ber
L_b & = & \frac{m_t}{f_{\tpi}\sqrt{2}}{\overline b}_L(H +iA^0)b_R +h.c \
\eer
where $f_{\tpi} \sim 50$ GeV is the top pion decay constant. On 
integrating out the Higgs fields $H$ and $A^0$ we have an effective 
four fermion operator
\ber
L_{eff} & = &\frac{m_t^2}{f_{\tpi}^2m_H^2}
{\overline b}_Lb_R 
{\overline b}_Rb_L  \
\eer
Since the $b$ quark in (4) is in 
the weak-eigenstate, $L_{eff}$ in (4) will induce
flavor changing neutral current (FCNC) four quark operators in Eq. (1) after 
diagonalizing the quark mass matrix\cite{burdman},
with coefficients,
\ber
R_1 & = & \frac{1}{4}\frac{m_t^2}{f_{\tpi}^2m_H^2}
|D_{Lbb}|^2D_{Rbb}D^*_{Rbs}\nonumber\\
R_2 &=& \frac{1}{4}\frac{m_t^2}{f_{\tpi}^2m_H^2}
|D_{Rbb}|^2D_{Lbb}D^*_{Lbs} \
\eer
where $D_L$ and $D_R$ are the mixing matrices in the left and the 
right handed down sector. In the charm sector similar interactions can arise 
due to the strong couplings of the top quark to top pions. 
The effective operators generated by integrating out the top-pions are similar to Eq. (1)
with the replacement of $b$ by $c$ and $s$ by $u$.
In topcolor II
models \cite{burdman,hill2}, where there can be strong top-pion
couplings of the top with the charm quark, we have 
\ber
R_1 & = & \frac{1}{4}\frac{m_t^2}{f_{\tpi}^2m_{\tpi}^2}
|U_{Lcc}|^2U_{Rtc}U^*_{Rtu}\nonumber\\
R_2 & = & \frac{1}{4}\frac{m_t^2}{f_{\tpi}^2m_{\tpi}^2}
|U_{Rtc}|^2U_{Lcu}U^*_{Lcc}\
\eer

\noindent
In supersymmetric standard models without $R$ parity,
the most general superpotential of 
the MSSM,
consistent with $SU(3)\times SU(2)\times U(1)$
 gauge symmetry and supersymmetry, can be written as
\begin{equation}
{\cal W}={\cal W}_R+{\cal W}_{\not \! R},
\end{equation}
where ${\cal W}_R$ is the $R$-parity conserving part while 
${\cal W}_{\not \! R}$ violates the $R$-parity. They are given by 
\begin{eqnarray}
{\cal W}_R&=&h_{ij}L_iH_2E_j^c+h_{ij}^{\prime}Q_iH_2D_j^c
             +h_{ij}^{\prime\prime}Q_iH_1U_j^c,\\ \label{RV}
{\cal W}_{\not \! R}&=&\lambda_{ijk}L_iL_jE_k^c
+\lambda_{ijk}^{\prime}L_iQ_jD_k^c
             +\lambda_{ijk}^{\prime\prime}U_i^cD_j^cD_k^c+\mu_iL_iH_2.
\end{eqnarray}
Here $L_i(Q_i)$ and $E_i(U_i,D_i)$ are the left-handed
lepton (quark) doublet and lepton (quark) singlet chiral superfields, with
$i,j,k$ being generation indices and $c$ denoting a charge conjugate field.
$H_{1,2}$ are the  chiral superfields
representing the two Higgs doublets. 
In the  $R$-parity violating superpotential above, the  
$\lambda$ and $\lambda^{\prime}$ couplings 
violate lepton-number conservation, while the
$\lambda^{\prime\prime}$ couplings violate baryon-number conservation.
$\lambda_{ijk}$ is antisymmetric in the first two
indices and $\lambda^{\prime\prime}_{ijk}$ is antisymmetric in
the last two indices.
While it is theoretically possible to have both baryon-number 
and lepton-number violating terms in the lagrangian, the non-observation
of proton decay imposes very stringent conditions on their simultaneous
presence \cite{proy}.
We, therefore, assume the existence of either  $L$-violating couplings or  
$B$-violating couplings, but not the coexistence of both. 
We calculate the effects of both types of couplings.

In terms of the four-component Dirac notation, the  lagrangian involving the
$\lambda^{\prime}$ and $\lambda^{\prime\prime}$ couplings  is given by
\begin{eqnarray}
{\cal L}_{\lambda^{\prime}}&=&-\lambda^{\prime}_{ijk}
\left [\tilde \nu^i_L\bar d^k_R d^j_L+\tilde d^j_L\bar d^k_R\nu^i_L
       +(\tilde d^k_R)^*(\bar \nu^i_L)^c d^j_L\right.\nonumber\\
& &\hspace{1cm} \left. -\tilde e^i_L\bar d^k_R u^j_L
       -\tilde u^j_L\bar d^k_R e^i_L
       -(\tilde d^k_R)^*(\bar e^i_L)^c u^j_L\right ]+h.c.,\\
{\cal L}_{\lambda^{\prime\prime}}&=&-\lambda^{\prime\prime}_{ijk}
\left [\tilde d^k_R(\bar u^i_L)^c d^j_L+\tilde d^j_R(\bar d^k_L)^c u^i_L
       +\tilde u^i_R(\bar d^j_L)^c d^k_L\right ]+h.c.
\end{eqnarray}
The terms proportional to $\lambda$ are not relevant to our present 
discussion and will not be considered here.
 The exchange of sneutrinos with the 
$\lambda^{\prime}$ coupling will generate $L_{new}$
for $\Upsilon \to {\overline B}X_s$ with
\ber
R_1 & = &\frac{1}{4}
 \Sigma_{i}\frac{\lambda^{\prime }_{i32} \lambda^{\prime *}_{i33}}
{m_{\tilde {\nu_i}}^2}\nonumber\\
R_2 &=&\frac{1}{4}
 \Sigma_{i}\frac{\lambda^{\prime *}_{i23} \lambda^{\prime}_{i33}}
{m_{\tilde {\nu_i}}^2}\
\eer
For the case of $J/\psi \rightarrow D X_u$ the operators in $L_{new}$ cannot be generated
at tree level.
\section{Results}
The various inputs to our calculations can be found in \cite{DOPZ}.
The standard model contribution 
to the branching ratio is
$5.2 \times 10^{-11}$ from the penguin induced $ b \to s$  transition.  The 
process $\Upsilon \to {\overline B} X_s$ can also have a contribution in the 
standard model from tree level processes. The effective Hamiltonian
, suppressing the Dirac structure of the currents, 
\ber
H_W & = & \frac{G_F}{\sqrt{2}}
        V_{ub}V_{us}^*\left[a_1({\overline u}b)({\overline s}u) +
a_2({\overline s}b)({\overline u}u)\right] \
\eer
where $a_1$ and $a_2$ are the QCD coefficients  can generate the process
$\Upsilon \to B^+ K^-$. 
We can estimate the branching ratio for this process as
$$ BR[\Upsilon \to B^+ K^-] \approx |\frac{V_{ub}}{V_{cb}}|^2
BR[\Upsilon \to B_c^+ K^-]$$
Using $ BR[\Upsilon \to B_c^+ K^-]$ calculated in Ref\cite{indianguys}
one obtains $BR[\Upsilon \to B^+ K^-] \sim 1.5 \times 10^{-14}$.
For a rough estimate of $BR[\Upsilon \to B^+ X_s]$ we can 
scale $BR[\Upsilon \to B^+ K^-]$ by the factor 
$BR[B \to D^0 X]/BR[B \to D \pi]$. 
The measured value of $BR[B \to D^0 X]$ 
\cite{PRD} includes $D^0$ coming from the decay of $D^{0*}$ and
$D^{+*}$. From the spin phase factors $BR[B \to D^* X] \sim 3 BR[B \to D X]$.
Hence $BR[B \to D^0 X]/BR[B \to D \pi] \sim 20$ leading to
$BR[\Upsilon \to B^+ X_s] \sim 3 \times 10^{-13}$.  
 
So far we have not considered $R_1$ and $R_2$ in $L_{new}$. 
 In our model independent analysis
we vary $R_1/\Lambda^2$, $R_2/\Lambda^2$ one at a time 
 and the use the 
constraint from measurements of $ b \to s e^+ e^{-}$ and
$ b \to s \mu^+ \mu^{-}$ . We 
 identify $\Lambda$ with the masses of the 
exchange particles which we take to be between $100-200$ GeV.
 The allowed values of
$R_1/\Lambda^2$, $R_2/\Lambda^2$  are then used to calculate
$\Upsilon \to {\overline B} X_s$
The constraint from $b \to s l^+ l^-$ gives
$$|R_{1,2}|/\Lambda^2 < (6-9)\times 10^{-6} (1/GeV)^2$$
Using the upper bounds on $|R_{1,2}|/\Lambda^2$ 
we find the branching ratio for the process  
$\Upsilon(1S) \to {\overline{B}} X_s$ 
to be between $(1-2) \times 10^{-6}$.  Branching ratios 
of similar order  are also obtained for
$\Upsilon(2S)$ and $\Upsilon(3S)$. For $\Upsilon(4S)$ the branching ratio is 
smaller by a factor of $100$ because of the larger width of
$\Upsilon(4S)$ which decays predominantly to two $B$ mesons.
 
Turning now to models, we find for the top color model from Eq. (5)
we can write
\ber
D^*_{Rbs} & = & 4\frac{R_1}{\Lambda^2}\frac{f_{\tpi}^2m_H^2}{m_t^2
|D_{Lbb}|^2D_{Rbb}}
\eer
We can identify $\Lambda=m_H$ and use the constraint from
$ b \to s e^+ e^{-}$ for a typical value of 
$|R_1|/\Lambda^2 \sim 6 \times 10^{-6} (1/GeV)^2$ .  
Assuming $|D_{Lbb}| \approx |D_{Rbb}| \approx 1$, and
$f_{\tpi}=50$GeV we obtain
$|D_{Rbs}| \sim 2 m_H^2 \times 10^{-6}$. With typical values of 
$m_H \sim 100-200$ GeV we get
$|D_{Rbs}| \sim 0.02-0.08$. Similar values have been obtained 
for $|D_{Rbs}|$ in Ref\cite{burdman} by considering the contributions of
 the charged higgs and top-pion  to $b \to s \gamma$. 
A similar exercise can be carried out with $|D_{Lbs}|$. Note
 that $B_s$ mixing probes
the combination $D^*_{Lbs}D_{Rbb}D^*_{Rbs}D_{Lbb}$ and so by either choosing
$R_1 \sim 0$ or $R_2 \sim 0$ we can satisfy the constraint on $B_s$mixing by 
choosing the appropriate mixing elements to be small. Note that in top color
models we can have operators 
${\overline s}(1-\gamma^5)b{\overline d} 
(1+\gamma^5) d $ and 
${\overline s}(1+\gamma^5)b{\overline d}
(1-\gamma^5) d $ that can contribute to
$ \Upsilon  \rightarrow {\overline B}   s {\overline d} \rightarrow
{\overline B} X_s $ 
after Fierz reordering. 
However these operators will be suppressed by form factor effects and also from mixing effects. We have checked that the contribution to
$ \Upsilon  \rightarrow {\overline B}  X_s $ from these operators are
much suppressed relative to the contribution of 
 the operators in $L_{new}$.  We will therefore not consider the the above operators in our analysis.

Turning to R-parity violating susy we first collect the 
constraints on the relevant couplings.
The upper limits of the $L$-violating couplings for the squark mass of
100 GeV are given by
\begin{eqnarray}
\vert \lambda^{\prime}_{kij}\vert&<&0.012,~(k,j=1,2,3; i=1,2),\\
\vert \lambda^{\prime}_{13j}\vert&<&0.16,~(j=1,2),\\
\vert \lambda^{\prime}_{133}\vert&<&0.001,\\
\vert \lambda^{\prime}_{23j}\vert&<&0.16,~(j=1,2,3),\\
\vert \lambda^{\prime}_{33j}\vert&<&0.26,~(j=1,2,3),
\end{eqnarray}
The first set of constraints in Eq. (15) come from 
 the decay $K\rightarrow \pi \nu
\nu$ with FCNC processes in the down quark sector
\cite{agashe}. The 
set of constraints in Eq. (16) and Eq. (18) are obtained 
from the semileptonic decays of 
$B$-meson \cite{29}. 
The  constraint,
 on the coupling $\lambda^{\prime}_{133}$ in Eq. (17)  is obtained from the 
Majorana mass that the coupling can generate for the 
electron type neutrino \cite{21}.
The last set of limits in Eq. (19)  are derived from the leptonic decay modes of 
the $Z$ \cite{30}.
Assuming all the couplings to be positive we find the branching ratio for 
$\Upsilon \to {\overline B} X_s$ to be around $2 \times 10^{-6}$
for $m_{\tilde {\nu}}=100$GeV.

Turning next to $J/\psi \to {\overline D} X_u$, we first make an 
estimate for this process in the standard model. Since the penguin 
$ c \to u$ transition is
small in the standard model we neglect its contribution. As in the 
case for the
$\Upsilon$ system , for a rough estimate,  can write
$$ BR[ J/\psi \to D^0 X_u] 
\sim BR[ J/\psi \to D^0 \pi^0]
BR[D^0 \to K^- X]/BR[D^0 \to K^- \pi^+]$$  We obtain
$BR[ J/\psi \to D^0 \pi^0]$ from \cite{indianguys} and keeping in mind that
$BR[D^0 \to K^- X]$ contains contributions from states decaying to $K^-$
 we obtain  
 $BR[ J/\psi \to D^0 X_u] \sim 10^{-10}$ . A similar exercise gives
 $BR[ J/\psi \to D^+ X_u] \sim 10^{-9}$.

Considering new physics effects we can constrain $R_1$ and $R_2$
from $ c \to u l^+ l^-$. We get an estimate of the
 constraint on 
$c \to u e^+e^{-}$ 
by adding up the exclusive modes
$$ BR[D \to u e^+e^{-}] \ge BR[ D \to (\pi^0 + 
\eta + \rho^0 +\omega) e^+e^{-}]$$ 
From $c \to u l^+ l^-$ one obtains
$$|R_{1,2}|/\Lambda^2 \le 3.7 \times 10^{-4} (1/GeV)^2$$
 We find the branching fraction for the process 
$J/\psi \to {\overline D} X_u $
using the constraint from
$ c \to u l^+ l^-$  can be
$(3-4)\times 10^{-5}$

 In top color models  taking $R_1$ and $R_2$ one at a time, one obtains
$$\frac{2.1 \times 10^3}{{m_{\tpi}^4}}||U_{Lcc}|^2U_{Rtc}U^*_{Rtu}|^2 $$
 or
$$\frac{2.1 \times 10^3}{m_{\tpi}^4} ||U_{Rtc}|^2U_{Lcc}U^*_{Lcu}|^2 $$
as  
the branching fraction 
for $J/\psi \to {\overline D} X_u$. For $m_{\tpi}$ between
$100-200$ GeV this rate can be between $(0.1-2.0) \times 10^{-5}$
if all the mixing angles are $\sim 1$.
It has been shown in 
Ref\cite{kominis2} that our choices for $f_{\tpi}$ and $m_{\tpi}$
gives unacceptably large corrections to $Z \to b{\overline b}$ from
one loop
contribution of the top pions. However in a strongly coupled theory higher
loop terms can have significant contributions. Nonetheless 
if  we change
 $f_{\tpi}$ to $\sim 100$ GeV for better agreement with
$Z \to b {\overline b}$ data then the effect in
$J/\psi \to {\overline D} X_u$  is reduced by a factor of 16. As in the case of the $\Upsilon$ system we can satisfy the constraint from D mixing by choosing
$R_1 \sim 0$ or $R_2 \sim 0$.
 
For R parity violating susy, contribution to $J/\psi \to {\overline D} X_u$
 can only occur at loop level,
 with both the $\lambda^{\prime}$ or $\lambda^{\prime \prime}$ contributing,
through the box diagram
 and so is suppressed.

 \section{Acknowledgements}
This work was done in collaboration with P. J. O'Donnell, S. Pakvasa and X. Zhang. Funding for the work was provided by
 Natural  Sciences and Engineering  Council
of Canada.

\end{document}